\documentstyle[12pt]{article}
\begin{document}
\renewcommand{\thesection}{\Roman{section}}
\renewcommand{\thesubsection}{\Alph{subsection}}
\renewcommand{\theequation}{\arabic{section}.\arabic{equation}}
\renewcommand{\thetable}{\arabic{section}.\arabic{table}}
\begin{titlepage}
\pagestyle{empty}
\baselineskip=21pt
\vskip .2in
\begin{center}
{\large{\bf
Momentum Distribution in Nuclear Matter within a Perturbation Approximation}}
 \end{center}
\vskip .1in

\begin{center}

A. Mariano$^{\dagger}$ and F. Krmpoti\'{c}$^{\dagger}$

{\small\it Departamento de F\'\i sica, Facultad de Ciencias Exactas,}\\
{\small\it Universidad Nacional de La Plata, C. C. 67, 1900 La Plata, Argentina}

{\small\it and}

A.F.R. de Toledo Piza

{\small\it Instituto de F\'\i sica, Universidade de S\~{a}o Paulo,\\
C.P. 20516, 01498 S\~{a}o Paulo, Brasil}

\end{center}
\vskip 0.5in
\baselineskip=18pt
It is shown that the norm corrections, introduced to avoid the violation
of the constraints on the depletion of the hole states in the standard 
perturbative $2p2h$ approach, leads in nuclear matter to a dependence of 
the momentum distribution with the total nucleon number. 
This unphysical behavior,
which in turn makes the depletion to be non-extensive, arises from
contributions of disconnected diagrams contained in the norm.
It is found that the extensivity is again recovered when the $4p4h$
excitations in the ground state are included, and a reasonable value
for the total number of nucleons promoted above the Fermi level is obtained.

\vspace{0.25in}
PACS number(s):
\bigskip

\vspace{0.5in}
\noindent
$^{\dagger}$Fellow of the CONICET from Argentina.
\end{titlepage}
 
\begin{center}
\section{INTRODUCTION}
\end{center}

A great amount of theoretical effort was devoted to study the influence of
ground-state correlations on the nuclear physical observables.
In most of them the nuclear response function is evaluated after introducing
two particle - two hole ($2p2h$) admixtures within first-order perturbation
into the $0p0h$ ground state wave function
\cite{Bertsch,Ericson,Arima,Nishizaki,Takayanagi1,Mariano1,Takayanagi2}.
It was found that $2p2h$ ground-state contribution modify significantly
the mean field results for the strength function. 
Recently, however, the use of this perturbation procedure has been objected
by Van Neck {\it et al.,} \cite{Neck} because it largely overestimates the effect of
ground-state correlation, when the norm corrections are neglected.
This happens because the perturbation adds a very large number of relatively
small excited $2p2h$ components, with relative weights in the perturbed
wavefunction typically large enough numerically as to
strongly enhance the norm corrections.
Specifically, it is reminded in Ref. \cite{Neck} that a shell-model approach
for occupation numbers in nuclei imposes constrains in the sense that
the number of particles lifted out the Fermi sea has to be $\leq 2$.
This constrain is not respected in the above mentioned works as put in
evidence  by several calculations of the occupation numbers within the
same perturbation approach\cite{TakaLipp}.

In a recent study \cite{Mariano2} of the Gamow-Teller strength the
ground-state  was worked out within a finite nuclei formalism up to the
first order, both in the Rayleigh-Schr\"odinger  and  Brioullin-Wigner
schemes, and including the effects of the  norm as was suggested in
Ref. \cite{Neck}. The result was the strong reduction of  $2p2h$ contribution
to the response function.
In the present work we intend to implement better controlled approximations
for the correlation structure of the ground-state for the case of nuclear
matter, which may have important consequences in the evaluation of the
response function in the quasielastic region. As a first step in this
direction we analyze here
the momentum distribution in the ground state. The reason for that is that,
as shown within a simple model by Takayanagi \cite{Takayanagi2},
the longitudinal response in the quasi free region is directly related
to the momentum distribution of nucleons in nuclear matter.
Our initial intention was to include the norm effects in the some way
as we have previously done for the finite nuclei \cite{Mariano2}.
But, we soon discovered that this procedure makes the depletion of the
nuclear core not to be anymore an extensive quantity because of the
momentum distribution dependence on the total nucleon number $A$.
Clearly, such a behavior is unphysical and has its origin in the
contributions arising from the infinite series of disconnected Goldstone
diagrams contained in the norm, as can be put in evidence by performing
a perturbation expansion of the later. In order to exactly cancel  these
disconnected graphs one is forced to enlarge the configuration space
by including $npnh$ amplitudes with $n>2$. This is done in the present
work for the $4p4h$ components in the ground state wave function.
To go beyond this configuration space is not only a cumbersome task
but hard to justify also.

\begin{center}
\section{FORMALISM}
\end{center}

The occupation number for the single-particle state $\kappa$ in the ground
state $|0\rangle$ is defined as
\begin{eqnarray}
n(\kappa)= \langle 0| a^{\dag}(\kappa)a(\kappa) |0\rangle,
\label{2.1}
\end{eqnarray}        
and the total number of particles promoted above the Fermi level, $N_>$,
and those remaining below the Fermi level, $N_<$, are
\begin{eqnarray}
N_>&=&  \sum_{\kappa (\epsilon_{_\kappa} > \epsilon_{_F})} n(\kappa),
\nonumber\\
N_< &=&\sum_{\kappa(\epsilon_{_\kappa} < \epsilon_{_F})} n(\kappa)= A -N_>.
\label{2.2} \end{eqnarray}
Within the Hartree-Fock (HF) approximation, $|0\rangle\equiv|HF\rangle =
|0p0h\rangle$ one obtains the well known step function for $n(\kappa)$, {\em i.e., }
$n(\kappa)= \theta(\epsilon_{_F}-\epsilon_{_\kappa})$, where
$\epsilon_{_\kappa}$ and $\epsilon_{_F}$ are, respectively, the single particle and Fermi energies.
This is the zero order approximation for the occupation numbers distribution.
When $npnh$ correlations are added to the ground state wave function,
states above the Fermi level are populated with the corresponding
depletion of the nuclear core, and the occupation number takes the form
\begin{eqnarray}
n(\kappa) = \theta(\epsilon_{_F} - \epsilon_{_\kappa})+\delta n(\kappa),
\label{2.3} \end{eqnarray}
being
\begin{equation}
N_> = \sum_{\kappa(\epsilon_{_\kappa} > \epsilon_{_F})}\delta n(\kappa) 
.\label{2.4}
\end{equation}

\begin{center}
\subsection{$2p2h$ correlations}
\end{center}

The ground state wave function, with the first order $2p2h$ perturbations
included to the HF state, reads
\begin{equation}
|0\rangle = {\cal N} \left[|HF\rangle+ \frac{1}{4}
\sum_{p's,h's} c_{p_1p_2h_1h_2} |p_1p_2h_1h_2\rangle\right],
\label{2.5}\end{equation}
where
\begin{eqnarray}              
c_{p_1p_2h_1h_2}=-\frac{\langle p_1p_2h_1h_2|\hat V |HF\rangle}
{E_{p_1p_2h_1h_2}},
\label{2.6} \end{eqnarray}
are the $2p2h$ amplitudes, and
\begin{eqnarray}              
{\cal N}  = \frac{1}{\sqrt{1 + {\cal N}_{2p2h}}},
\label{2.7} \end{eqnarray}
is the overall normalization factor, with
\begin{eqnarray}              
{\cal N}_{2p2h} \equiv \frac{1}{4}\sum_{p's,h's}
\left|c_{p_1p_2h_1h_2}\right|^2.
\label{2.8} \end{eqnarray}
$|p_1p_2h_1h_2\rangle$ and $E_{p_1p_2h_1h_2}$ stand, respectively, for
the $2p2h$ configurations and single-particle energies, and $\hat{V}$
is the residual interaction.

The usual second-order approximation for $n(\kappa)$ is obtained by
retaining only the first term in the expansion
\begin{eqnarray}
{\cal N}^2=1-{\cal N}_{2p2h}+ {\cal N}_{2p2h}^2-\cdots,
\label{2.9} \end{eqnarray}
and for $\delta n(\kappa)$ one gets
\begin{eqnarray}
\delta n^{(2)}(\kappa) =  \frac{1}{2} \sum_{p's,h's}
(\delta_{\kappa,p_1} - \delta_{\kappa,h_1})
\left[\frac{ V_{p_1 p_2 h_1h_2}}
{E_{p_1p_2h_1h_2}}\right]^2,
\label{2.10} \end{eqnarray}
where $V_{p_1 p_2 h_1h_2}$ stand for the antisymmetrized matrix elements of
$\hat{V}$.
The contribution of $\delta n(\kappa)^{(2)}$ is schematically shown in
Fig. \ref{fig1}, and
\begin{equation}
N_>^{(2)}= 2{\cal N}_{2p2h}.
\label{2.11} \end{equation}
This result is only valid
when ${\cal N}_{2p2h}\ll1$, condition which is not fulfilled in most of
the numerical calculations performed so far. This was precisely the reason
why Van Neck {\it et al.,} \cite{Neck} have proposed to use the second order normalized 
approximation (with ${\cal N}$  given by (\ref{2.7})). This leads to
the variation in the occupation numbers 
\begin{equation}
\delta n^{(2N)}(\kappa) = {\cal N}^2 \delta n^{(2)}(\kappa)
\label{2.12} \end{equation}
and the corresponding number of promoted particles is 
\begin{equation}
N^{(2N)}_> = {\cal N}^2 N_>^{(2)}
=2\frac{{\cal N}_{2p2h}}{1+{\cal N}_{2p2h}} \leq 2,
\label{2.13} \end{equation}
which is a satisfactory bound for $N_>$. Yet, $\delta n^{(2N)}(\kappa)$ contains
an infinite series of \linebreak disconnected diagrams, illustrated in Fig. 
\ref{fig2}. 
We show below how they are removed by enlarging the configuration space.

\begin{center}
\subsection{$4p4h$ correlations}
\end{center}

The ground-state wavefunction
with the $2p2h$ and $4p4h$ admixtures included is written as
\begin{eqnarray}
|0\rangle & = & {\cal N} \left[|HF \rangle + \frac{1}{(2!)^2}
\sum_{p's,h's}
c_{p_1p_2h_1h_2} |p_1p_2h_1h_2\rangle \right .\nonumber \\
& + & \left .\frac{1}{(4!)^2}
\sum_{p's,h's}
c_{p_1p_2p_3p_4h_1h_2h_3h_4} |p_1p_2p_3p_4h_1h_2h_3h_4\rangle \right  ], 
\label{2.14}
\end{eqnarray}        
where
\begin{eqnarray}              
c_{p_1p_2p_3p_4h_1h_2h_3h_4}=\frac{\langle HF |\hat V | p_1p_2h_1h_2 \rangle
\langle p_1p_2h_1h_2|\hat V |  p_1p_2p_3p_4h_1h_2h_3h_4 \rangle }
{E_{p_1p_2h_1h_2}E_{p_1p_2p_3p_4h_1h_2h_3h_4}},
\label{2.15} \end{eqnarray}
and
\begin{eqnarray}              
{\cal N}  = \frac{1}{\sqrt{1+{\cal N}_{2p2h}+{\cal N}_{4p4h}}},
\label{2.16} \end{eqnarray}
with 
\begin{equation}
{\cal N}_{4p4h}=\frac{1}{(4!)^2}
\sum_{p's,h's}\left|c_{p_1p_2p_3p_4h_1h_2h_3h_4}\right|^2,
\label{2.17}
\end{equation}
which is the norm factor coming from the $4p4h$ contributions.

The occupation numbers in the ground state (\ref{2.14}) read
\begin{equation}
\delta n(\kappa) = 
{\cal N}^2~ \left[\delta n^{(2)}(\kappa)+\delta n^{(4)}(\kappa)\right],
\label{2.18} \end{equation}
where
\begin{eqnarray}
\delta n^{(4)}(\kappa) & = & 
\frac{4}{(4!)^2}\sum_{p's,h's}(\delta_{\kappa,p_1} - \delta_{\kappa,h_1})
 \left|c_{p_1p_2p_3p_4h_1h_2h_3h_4}\right|^2.
\end{eqnarray}
\label{2.19}
Now, making an expansion up to the fourth order
\begin{equation}
\delta n(\kappa) = 
[1-{\cal N}_{2p2h}]~\delta n^{(2)}(\kappa)+\delta n^{(4)}(\kappa),
\label{2.20} \end{equation}
and after working out the $4p4h$ amplitudes we get 
\begin{equation}
\delta n^{(4)}(\kappa) = 
{\cal N}_{2p2h}\delta n^{(2)}(\kappa)  + \delta n^{(4C)}(\kappa)
,\label{2.21}
\end{equation}
or
\begin{equation}
\delta n(\kappa) = \delta n^{(2)}(\kappa)  + \delta n^{(4C)}(\kappa)
,\label{2.22}
\end{equation}
with
\begin{eqnarray}
& & \hspace{-17mm}\delta n^{(4C)}(\kappa) =  \frac{1}{4}\sum_{p's,h's}
\frac{ V_{p_1p_2 h_1h_2}V_{p_3p_4 h_3h_4}}{E_{p_1p_2h_1h_2}E_{p_3p_4h_3h_4}} 
\nonumber\\
& &\times\left\{-(\delta_{\kappa,p_1}+
\delta_{\kappa,p_2} - \delta_{\kappa,h_1} 
 - \delta_{\kappa,h_2})\left[
\frac{V_{p_1 p_2 h_1h_3}V_{p_3p_4 h_2h_4}}{E_{p_1p_2h_1h_3}E_{p_3p_4h_2h_4}}
+
\frac{V_{p_1 p_3 h_1h_2} V_{p_2p_4 h_3h_4}}{E_{p_1p_3h_1h_2} E_{p_2p_4h_3h_4}} 
\right] \right. \nonumber \\
& &\hspace{7mm}+ \left. (\delta_{\kappa,p_1} - \delta_{\kappa,h_1})
\left[4 \frac{ V_{p_1 p_3 h_1h_3} V_{p_2p_4 h_2h_4}}
{E_{p_1p_3h_1h_3}E_{p_2p_4h_2h_4}}
+\frac{1}{2}
\frac{V_{p p_2 h_3h_4}V_{p_3p_4h_1h_2}}{E_{pp_2h_3h_4} E_{p_3p_4h_1h_2}}
\right] \right\},
\label{2.23} \end{eqnarray}
being the contribution of the connected diagrams illustrated in 
Fig. \ref{fig3}.
The diagrams (a), (b) and (c) describe the different contributions that
can arise from the first term, while the graph (d) represents the second term. 
For the depletion number we now get
\begin{eqnarray}
N_> & = & 2{\cal N}_{2p2h} - 2{\cal N}_{2p2h}^2 
+ 4{\cal N}_{4p4h} =  2{\cal N}_{2p2h} + 
4{\cal N}_{4p4h}^C, \label{2.24}
\end{eqnarray}
where the last term comes from $\delta n^{(4C)}(\kappa)$. 
Note that the result (\ref{2.24})  corresponds to the  expansion up to 
fourth order of the quantity
\begin{equation}
\frac{2{\cal N}_{2p2h} + 4{\cal N}_{4p4h}}
{1+{\cal N}_{2p2h}+{\cal N}_{4p4h}} \leq 4 \label{2.25}
\end{equation}
which is the total depletion number obtained from the exact occupation 
probabilities (\ref{2.18}), and same as (\ref{2.13}), is also properly bounded.

Regarding the results of this subsection, it should be remarked that:
\begin{enumerate}
\item [(a)] In the expansion (\ref{2.9}) are retained the first two terms.
\item [(b)] The contributions to $\delta n(\kappa)$ coming from the 
disconnected graphs brought around by ${\cal N}_{2p2h}$ are not present in 
final result  because they cancel out with topological similar contributions 
arising in the wave function (\ref{2.14}), {\em i.e., } the first term of 
$\delta n^{(4)}(\kappa)$ in Eq.(\ref{2.21}).
\item [(c)] The norm term ${\cal N}_{4p4h}$ does not contribute 
in a calculation up to the fourth order and 
has been ignored in the expansion of ${\cal N}^2$. 
\item [(d)]
We do not include the coupling among the $2p2h$
states ($\langle 2p2h|\hat V|2p2h'\rangle = 0$), because it would only lead 
to a redistribution of the $2p2h$ occupation probabilities, which does not
have any effect on  $\delta n^{(2)}(k)$.  
\end{enumerate}

\begin{center}
\subsection{Evaluation of the occupation numbers distribution in
nuclear matter}
\end{center}

In nuclear matter the HF ground state is approximated by the Fermi gas,
the single particle quantum numbers $\kappa$ are the momentum ${\bf k}$,
the spin projection $m_s$ and the isospin projection $m_t$ of the particle,
and $n(\kappa)$ turns into the momentum distribution
\begin{eqnarray}
n(k) = \theta(1-k)+\delta n(k),
\label{2.26} \end{eqnarray}
where $k$ is measured in units of the wave number $k_F$ that defines the Fermi surface.
The corresponding depletion number now reads
\begin{eqnarray}
N_>= 3 A \int_1^{\infty} d k k^2 \theta(k - 1)\delta n(k).
\label{2.27} \end{eqnarray}
The residual interaction is generically expressed  in the form \cite{Mariano1}
\begin{eqnarray}
\hat{V}({\bf q}) = \sum_IV^I(q){\bf O}^I_1(\hat{\bf q})\cdot {\bf O}^I_2(-\hat{\bf q})
\label{2.28} \end{eqnarray}
where $V^I(q)$ are the interaction strengths 
and
$I\equiv T,S,J$ encompasses the isospin, spin and total angular momentum
quantum numbers of the operators ${\bf O}^I(\hat{\bf q})$, defined as
\begin{eqnarray}
{\bf O}^{000}(\hat{\bf q})&=&1;\hspace{0.5cm}~{\bf O}^{010}(\hat{\bf q})=
i(\hat{\bf q}\cdot\mbox{\boldmath$\sigma$});\hspace{0.5cm}{\bf O}^{011}(\hat{\bf q})=
(\hat{\bf q}\times\mbox{\boldmath$\sigma$}),\nonumber \\
{\bf O}^{100}(\hat{\bf q})&=&\mbox{\boldmath$\tau$};\hspace{0.5cm}{\bf O}^{110}(\hat{\bf q})=
i(\hat{\bf q}\cdot\mbox{\boldmath$\sigma$})\mbox{\boldmath$\tau$}\hspace{0.5cm}
{\bf O}^{111}(\hat{\bf q})=(\hat{\bf q}\times\mbox{\boldmath$\sigma$})\mbox{\boldmath$\tau$}.
\label{2.29} \end{eqnarray}
The exchange contributions to $\delta n(k)$ will be dropped out since,
as pointed out Van Order and Donnelly \cite{Donnelly}, they are small in
comparison with the direct ones.
In this way we get
\begin{eqnarray}
\delta n^{(2)}(k)&=&
\frac{1}{4} \int d{\bf q}
\left[\theta(k-1)\theta(1-|{\bf k} + {\bf q}|)
{\cal F}({\bf k}\cdot{\bf q} + q^2,q)\right.\nonumber\\
& + & \left.\theta(1-k)~\theta(|{\bf k} - {\bf q}|-1)
{\cal F}({\bf k}\cdot{\bf q} ,q)\right]
\sum_I [v^I(q)]^2 (2T+1) 2^J,
\label{2.30} \end{eqnarray}
for the second order correction to $n(k)$,
\begin{eqnarray}
{\cal N}_{2p2h} = \frac{3A}{8 \pi} \int d{\bf k} d{\bf q} \theta(k-1)
\theta(1-|{\bf k} + {\bf q}|){\cal F}({\bf k}\cdot{\bf q} + q^2,q) \sum_I [v^I(q)]^2
(2T+1) 2^{J},
\label{2.31} \end{eqnarray}
for second order correction to the norm, and
\begin{eqnarray}
\delta n^{(4C)}(k)
& = & - \frac{1}{16} \int d{\bf l} \int d{\bf p} \int d{\bf q} 
\theta(1- |{\bf l} + {\bf q}|)
\theta( {\bf t}- 1)\nonumber \\
&\times& [\theta(k-1){\cal G}({\bf k},{\bf l},{\bf p},{\bf q}) +
\theta(1-k){\cal G}({\bf k}-{\bf q},{\bf l},{\bf p},{\bf q})], 
\label{2.32}
\end{eqnarray}
for the fourth order correction to $n(k)$. In the above equations
\begin{equation}
v^I(q) = \frac{2 k_{_F}^3}{(2\pi)^3 \epsilon_F } V^I(q),
\label{2.33}\end{equation}
\begin{eqnarray}
{\cal F}(\alpha,q)&=&\int\frac{\theta(x -1)
\theta(1-|{\bf x} + {\bf q}|)}{(\alpha +  {\bf q}\cdot{\bf x} )^2}d{\bf x}\nonumber\\
& = & \frac{2\pi}{q}\left\{\left[-1+\ln 2+ \ln\left|
\frac{q^2-\alpha+q}{q^2-2\alpha}\right|
-\frac{\alpha}{q^2} \ln\left| \frac{q^2-\alpha+q}{q-\alpha}\right| \right]
\theta(2-q) \right.  \nonumber \\
&+&\left.\left[\ln\left| \frac{q^2-\alpha+q}{q^2-\alpha-q}\right|
-\frac{\alpha}{q^2} \frac{q^2-\alpha+q}{q^2-\alpha-q}+ \frac{2}{q}\right]
\theta(q-2) \right \},
\label{2.34} \end{eqnarray}
and
\begin{eqnarray}
&&{\cal G}({\bf k},{\bf l},{\bf p},{\bf q})=
\left[\theta(1-|{\bf l}+{\bf p}|){\cal F}({\bf l}\cdot{\bf p}+l^2,l)+
\theta(1-|{\bf k}+{\bf p}|){\cal F}({\bf k}\cdot{\bf p}+l^2,l)\right.\nonumber\\
&+&\left.\theta(|{\bf l}+{\bf q}+{\bf p}|-1){\cal F}(-{\bf l}\cdot{\bf p}-{\bf q}\cdot{\bf p},l)
+\theta(|{\bf k}+{\bf q}+{\bf p}|-1){\cal F}(-{\bf k}\cdot{\bf p}- {\bf q}\cdot{\bf p},l)\right]
\nonumber \\
&\times&\frac{\theta(1-|{\bf k}+{\bf q}|)}{({\bf k}\cdot{\bf q}+q^2+{\bf l}\cdot{\bf q})^2}
\sum_I[v^I(q)]^2(2T+1)2^{J}\sum_{I'}[v^{I'}(l)]^2(2T'+1)2^{J'}.
\label{2.35}  \end{eqnarray}
In Eq. (\ref{2.32}) we do not include the contributions
arising from the second term in Eq. (\ref{2.23}), because, as discussed
later on, they are relatively small.

From Eqs. (\ref{2.11}), (\ref{2.13}) and (\ref{2.31}) it can be easily seen that
$N^{(2)}_>$ fulfills the extensively condition, but
$N_>^{(2N)}(k)$ does not.
This  unphysical behavior of $N^{(2N)}_>$ is due to the disconnected
diagrams contained in $\delta n^{(2N)}(k)$
and is circumvented by including the $4p4h$ excitations. This lead us to
the result in Eq. (\ref{2.24}), which again satisfies the requirement on
extensivity.
\setcounter{equation}{0}
\begin{center}
\section{NUMERICAL RESULTS AND CONCLUSIONS}
\end{center}
The momentum  distribution $n(k)$ has been evaluated for the normal nuclear
matter density, {\it i.e.,} $k_{_F}=1.36$ fm, and by parameterizing the
residual interaction (\ref{2.28}) as follows:
\begin{eqnarray}
&  & 
V^{000}(q) = C_{\pi}(q) f,~~~ V^{100}(q) = 
C_{\pi}(q) f',~~~ V^{010}(q) = V^{011} (q)= C_{\pi}(q) g, \nonumber \\
\vspace{5mm} 
&  & 
V^{110}(q) = 
C_{\pi}(q)\left( g' - \frac{q^2}{q^2 + m_{\pi}^2} \right ),~~~ 
V^{111}(q) = C_{\pi}(q) g',  \label{2.36}
\end{eqnarray}
Here $f,~f',~g$ and $g'$ are the Landau-Migdal parameters,
$C_{\pi}(q)= f_{\pi}^2/\mu_{\pi}^2 \Gamma_{\pi}^2(q)$,
$m_{\pi} = \mu_{\pi}/k_{_F}$ and
\begin{eqnarray}
\Gamma_{\pi}(q)=\frac{\Lambda_{\pi}^2 - m_{\pi}^2 } {\Lambda_{\pi}^2 + q^2 },
\label{2.37} \end{eqnarray}
is the $\pi NN$ form factor.

We do not incorporate explicitly the $\rho$ meson since, as pointed out by
Dickhoff \cite{Dick1},  it produces too much suppression of the tensor force.
Yet, the contribution of heavier mesons is taken into account  empirically
by fixing the values of $g'$ and $\Lambda_{\pi}$ (as done by Alberico
{\it et al.,} \cite{Ericson}) to reproduce the effective tensor and central
components of the force derived by Dickhoff from a G-matrix calculation
(\cite{Dick1,Dick2}). This procedure yields $g'=0.5$ and
$\Lambda_{\pi}=800~MeV/(\hbar c k_{_F}) $.
On the other hand, for $f$, $f'$ and $g$ we have found more
appropriate to adopt the values obtained by B\"{a}ckman {\it et. al.} \cite{Back}
from the analysis of stability conditions and empirical values of
physical observables, {\it i.e.,} $f=-0.12$, $f'=0.39$ and $g = 0$.

The following three different ways of adding the ground state correlations
to the Fermi gas distribution will be numerically compared:
\begin{enumerate}
\item The usual $2p2h$ second order approximation,
{\it i.e., } Eq. (\ref{2.30}) is employed in the evaluation of $\delta n(k)$.
\item The above approach is modified by including the effect of the norm
through Eqs. (\ref{2.12}), (\ref{2.7}) and (\ref{2.31}).
\item Fourth order approximation within the $2p2h+4p4h$
subspace, in which $\delta n(k) $ is given by Eqs. (\ref{2.22}) and (\ref{2.32}).
\end{enumerate}

The momentum space integrals have been evaluated for energies up to $300~MeV$.
This limit has been established by examining the behavior of $N_>/A$
(with $\delta n(k)$ given by (\ref{2.30})) as function of the cutoff energy.
The results are displayed in Fig. \ref{fig4}.

The resulting momentum distributions $n(k)$ are shown in Fig.
\ref{fig5}. In the approximation 2), where the results are $A$ dependent,
we have chosen, as an example, the $^{40}Ca$ nucleus.
For this case the empirical occupation numbers are available from an optical
potential analysis of experimental cross sections \cite{Joh}
(see also ref. \cite{Nishizaki}).
The second order approach produces a pronounced
high-momentum tail above the Fermi level with a corresponding strong reduction
of the low-momentum part as compared with the Fermi gas. In the normalized
case 2), the previous variation of momentum distribution $\delta n(k)$ is just
rescaled by the factor ${\cal N}^2=0.11$ and this diminishes drastically the
effect of the ground state correlations. (Note that for $A=90$
${\cal N}^2=0.05$.) Within the approximation 3) the depletion of the momentum
distribution is quite significant, but still relatively small in comparison
with that obtained in the plain $2p2h$ approach.
This effect can be interpreted as the reduction of the $2p2h$ occupation
probabilities by the coupling with $4p4h$ states. It is worth noting that
the effect of the last term in Eq. (\ref{2.23}) cannot be distinguished
visually in Fig. \ref{fig5}.

The calculated numbers of particles shifted above the Fermi level, for the
implemented approximations, turn out to be $N_>=16.8,~1.8$ and $1.9$,
respectively. The last result is, not only physically sound, but also
consistent with the above mentioned
study \cite{Joh}, which yields $N_>(^{40}Ca)\cong 2.7$.

We summarize our conclusions as follows:
\begin{itemize}
\item The plain second-order approximation for the momentum distribution
grossly overestimates the effect of the $2p2h$ ground-state correlation.
This is put in evidence by the very large result obtained for the number
of particles shifted above the Fermi level, as already
pointed out \cite{Neck}.

\item The introduction of norm corrections, following the recipe given in
Ref. \cite{Neck}, leads unavoidably to an $A$ dependent momentum distribution,
which is engendered by the disconnected diagrams present in the norm.
As a consequence the number depletion becomes a non-extensive quantity
for the nuclear matter, which is very strong drawback of the approximation.

\item We have found out that the just mention unphysical contributions of the
disconnected diagrams are cancelled by the inclusion of $4p4h$
correlations. Besides, the contribution of the $4p4h$ connected  diagrams
strongly hinder the effect of $2p2h$ correlations, yielding a momentum
distribution with properly bounded depletion number.

\item The present results seem to indicate that the $2p2h$
and $4p4h$ correlations are the dominant degrees of freedom to be considered in the description of the nuclear ground state. In addition,
we feel that the evaluation of the momentum distribution up to fourth order
is a reasonably good approximation and that is not crucial the inclusion
of higher order correlations beyond the $4p4h$ ones.
\end{itemize}

\bigskip

This work was supported in part by a Fundaci\'on Antorchas grant.\\

\newpage
\begin{figure}
\begin{center} { \bf Figure Captions} \end{center}

\caption{ \protect \footnotesize
Goldstone diagrams corresponding to the second-order correction 
$\delta n^{(2)}(\kappa)$. Each line indicates schematically a particle
or a hole state, the dots represent the residual interaction and
the encircled dots correspond to the number operator.
\label{fig1}}

\caption{ \protect \footnotesize 
Goldstone diagrams  corresponding to the expansion of the norm in the
normalized second-order $2p2h$ approximation $\delta n^{(2N)}(\kappa)$.
\label{fig2}}

\caption{ \protect \footnotesize
Goldstone diagrams corresponding to the different contributions
in the fourth order $4p4h$ correction  $\delta n^{(4C)}(\kappa)$.
\label{fig3}}

\caption{ \protect \footnotesize
Plot of the second-order $2p2h$ approximation for $N_>/A$, as function
of the cutoff energy $E_c$ (measured in MeV) for  $2p2h$ excitations.
\label{fig4}}

\caption{ \protect \footnotesize
Momentum distribution $n(\kappa)$ for the different
approximations: Fermi gas (thin full lines), second-order approximation
with $2p2h$ correlations ($\delta n(\kappa) = \delta n^{(2)}(\kappa)$)
(long-dashed lines),
normalized second-order approximation for $^{40}Ca$
($\delta n(\kappa) = \delta n^{(2N)}(\kappa)$)
(short-dashed lines), and
fourth-order approximation with $2p2h+4p4h$ correlations,
($\delta n(\kappa) = \delta n^{(2)}(\kappa) + \delta n^{(4C)}(\kappa)$) (thick full lines).
\label{fig5}}
\end{figure}

\newpage
\pagestyle{empty}
\begin{picture}(200,300)(180,200)
\setlength{\unitlength}{1.4pt}
\end{picture}

\newpage
\pagestyle{empty}
\begin{picture}(200,300)(180,200)
\setlength{\unitlength}{1.4pt}
\thicklines
\put (245,300){\oval (30,70)}
\put (245,300){\oval (16,70)}
\put (245,335){\circle*{6}}
\put (245,265){\circle*{6}}     
\put (230,300){\circle{6}} 
\put (230,300){\circle*{3}} 
\put (230,235){{\huge Fig. 1}}
\put (360,140){\oval (30,70)}
\put (360,140){\oval (16,70)}
\put (360,175){\circle*{6}}
\put (360,105){\circle*{6}}     
\put (345,140){\circle{6}} 
\put (345,140){\circle*{3}} 

\put (195,140){\oval (30,50)}
\put (195,140){\oval (16,50)}
\put (195,165){\circle*{6}}
\put (195,115){\circle*{6}}     
\multiput (250,140)(40,0){2}{\oval (30,50)}
\multiput (250,140)(40,0){2}{\oval (16,50)}
\multiput (250,165)(40,0){2}{\circle*{6}}
\multiput (250,115)(40,0){2}{\circle*{6}}     
\put (145,140){{\Huge $[$}}
\put (333,140){{\Huge $]$}}
\put (150,140){{\Large \bf $~1~ -$}}
\put (217,140){{\Large \bf $+$}}
\put (310,140){{\Large \bf  $.~.~.$}}
\put (230,80){{\huge Fig. 2}}
%
%
%
%
\end{picture}
\newpage
\begin{picture}(200,300)(105,50)
\setlength{\unitlength}{1.4pt}
\thicklines

\multiput (100,150)(80,0){4}{\oval (30,50)[r]}
\multiput (100,150)(80,0){4}{\oval (16,50)[r]}
\multiput (100,175)(80,0){3}{\line (0,-1){50}}
\multiput (100,175)(80,0){4}{\circle*{6}}
\multiput (100,125)(80,0){4}{\circle*{6}}     
\multiput (70,185)(80,0){3}{\line (1,-2){30}}
\multiput (70,115)(80,0){3}{\line (1, 2){30}}
\multiput (70,185)(80,0){3}{\line (0,-1){70}} 
\multiput (70,150)(80,0){4}{\oval (30,70)[l]}
\multiput (70,150)(80,0){4}{\oval (16,70)[l]}
\multiput (70,185)(80,0){4}{\circle*{6}}
\multiput (70,115)(80,0){4}{\circle*{6}}     
\put (55,150){\circle{6}}               
\put (55,150){\circle*{3}}               
\put (150,150){\circle{6}}               
\put (150,150){\circle*{3}}               
\put (242,160){\circle{6}}               
\put (242,160){\circle*{3}}               
\put (85,90){{\huge(a)}}
\put (165,90){{\huge (b)}}
\put (245,90){{\huge (c)}}
\put (325,90){{\huge (d)}}
\put (180,55){{\huge Fig. 3}}
\put (295,150){\circle{6}}               
\put (295,150){\circle*{3}}               
\put (308,185){\line (1,-2){30}}
\put (308,115){\line (1, 2){30}}
\put (312,185){\line (1,-2){30}}
\put (312,115){\line (1, 2){30}}

\end{picture}

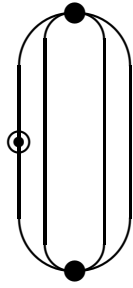
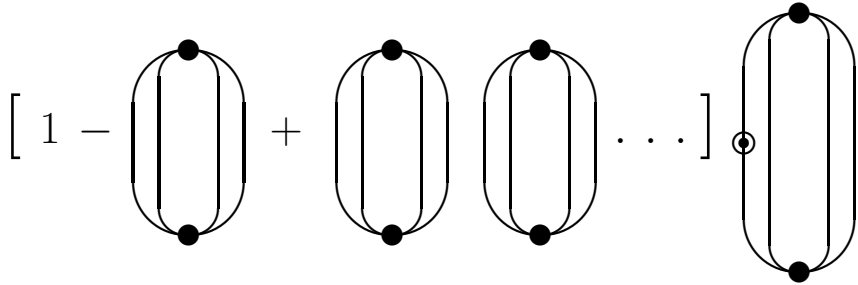
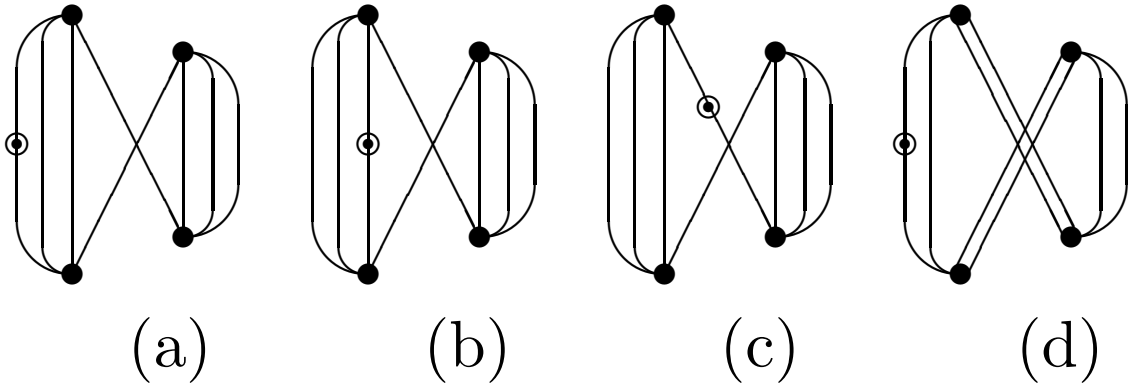
\end{document}